%% file: LKCell.tex
\documentclass[review]{elsarticle}

\usepackage{lineno}
\modulolinenumbers[5]

\journal{Neurocomputing}

\usepackage{amsmath,amsfonts}
\usepackage{algorithm}
\usepackage{array}
\usepackage[caption=false,font=normalsize,labelfont=sf,textfont=sf]{subfig}
\usepackage{textcomp}
\usepackage{stfloats}
\usepackage{url}
\usepackage{verbatim}
\usepackage{adjustbox}
\usepackage[colorlinks,linkcolor=red]{hyperref}
\usepackage{doi}
\usepackage{amssymb}
\usepackage{booktabs}
\usepackage[numbers]{natbib}
\usepackage{algpseudocode}
\usepackage{rotating}
\usepackage{xcolor}
\usepackage{graphicx}
\usepackage{subcaption}

\usepackage{multirow}

\begin{document}

\begin{frontmatter}

\title{LKCell: Efficient Cell Nuclei Instance Segmentation with Large Convolution Kernels}

\author[hustei]{Ziwei Cui\corref{equal}}
\author[hustei]{Jingfeng Yao\corref{equal}}
\author[hustei]{Lunbin Zeng}
\author[hg]{Juan Yang}
\author[hustei]{Wenyu Liu}
\author[hustei]{Xinggang Wang\corref{corr}}\ead{xgwang@hust.edu.cn} 

\cortext[equal]{Equal Contribution}
\cortext[corr]{Corresponding author}
\address[hustei]{School of Electronic Information and Communications, Huazhong University of Science and Technology, Wuhan 430074, Hubei Province, China}
\address[hg]{Department of Cardiology, Huanggang Central Hospital, Huanggang 438000, Hubei Province, China}

\begin{abstract}
The segmentation of cell nuclei in tissue images stained with the blood dye hematoxylin and eosin (H$\&$E) is essential for various clinical applications and analyses. Due to the complex characteristics of cellular morphology, a large receptive field is considered crucial for generating high-quality segmentation. 
However, previous methods face challenges in achieving a balance between the receptive field and computational burden.
To address this issue, we propose LKCell, a high-accuracy and efficient cell segmentation method. Its core insight lies in unleashing the potential of large convolution kernels to achieve computationally efficient large receptive fields. Specifically, 
(1) We transfer pre-trained large convolution kernel models to the medical domain for the first time, demonstrating their effectiveness in cell segmentation.
(2) We analyze the redundancy of previous methods and design a new segmentation decoder based on large convolution kernels. It achieves higher performance while significantly reducing the number of parameters.
We evaluate our method on the most challenging benchmark and achieve state-of-the-art results (0.5080 mPQ) in cell nuclei instance segmentation with only 21.6\% FLOPs compared with the previous leading method. Our source code and models are available at \url{https://github.com/hustvl/LKCell}.
\end{abstract}

\begin{keyword}
Nuclei Segmentation \sep Instance Segmentation \sep Large Kernel
\end{keyword}

\end{frontmatter}


\section{Introduction}
\input{intro}

\section{Related Work}
\input{related}

\section{Method}
\input{methods}

\section{Experiment}
\input{exp}

\section{Conclusion}
Cell nucleus instance segmentation is crucial in clinical applications, requiring reliable and automated segmentation models. In this paper, we propose a novel U-net-shaped cell nucleus segmentation network with large convolution kernels. We demonstrate state-of-the-art performance in cell nucleus instance segmentation on the PanNuke dataset, achieving the best results with minimal computational requirements. Furthermore, the generalization ability of our model is evident in the MoNuSeg dataset as a test dataset. The combination of low FLOPs and superior performance provides our model with a significant advantage for future clinical applications.

\bibliography{sample-base}

 \newpage

\vfill

\end{document}

%% file: intro.tex

Cell physiology and pathology analysis play a crucial role in clinical diagnosis and treatment. In cancer diagnosis and treatment, parameters like tumor cell density, nucleus-to-cytoplasm ratio, and average cell size are vital for cancer grading and prognosis \cite{castleman1996digital}. Recently, with the rapid development of deep learning, cell segmentation methods based on deep learning have emerged~\cite{horst2023histology,lu2021data,ester2023valuing}. They automatically segment given cell images, reducing the burden on the healthcare system. 

However, achieving high-performance and efficient cell segmentation is still challenging due to problems like uneven staining, cell overlap, and cluster morphology~\cite{ilyas2022tsfd}. 
Previous methods~\cite{chen2023cpp, ilyas2022tsfd, schmidt2018cell} achieve automatic cell segmentation with stacking convolution layers (mostly with 3$\times$3 kernels) and U-shape architecture~\cite{ronneberger2015u}. 
These methods are simple and efficient, yet their performance is not satisfactory due to the limitations of the receptive field. Recently, the Vision Transformer (ViT)~\cite{dosovitskiy2021image} has introduced new possibilities to medical segmentation~\cite{na2024segment,xu2023sppnet} with its powerful modeling capabilities and global receptive field. The most advanced cell segmentation method~\cite{horst2024cellvit} achieves state-of-the-art performance by incorporating a pre-trained, large parameter ViT backbone~\cite{SAM}, demonstrating the importance of receptive field in cell segmentation. Nevertheless, this also results in high computational costs, severely limiting its widespread application in clinical settings.


\begin{figure}
    \centering
    \includegraphics[width=\linewidth]{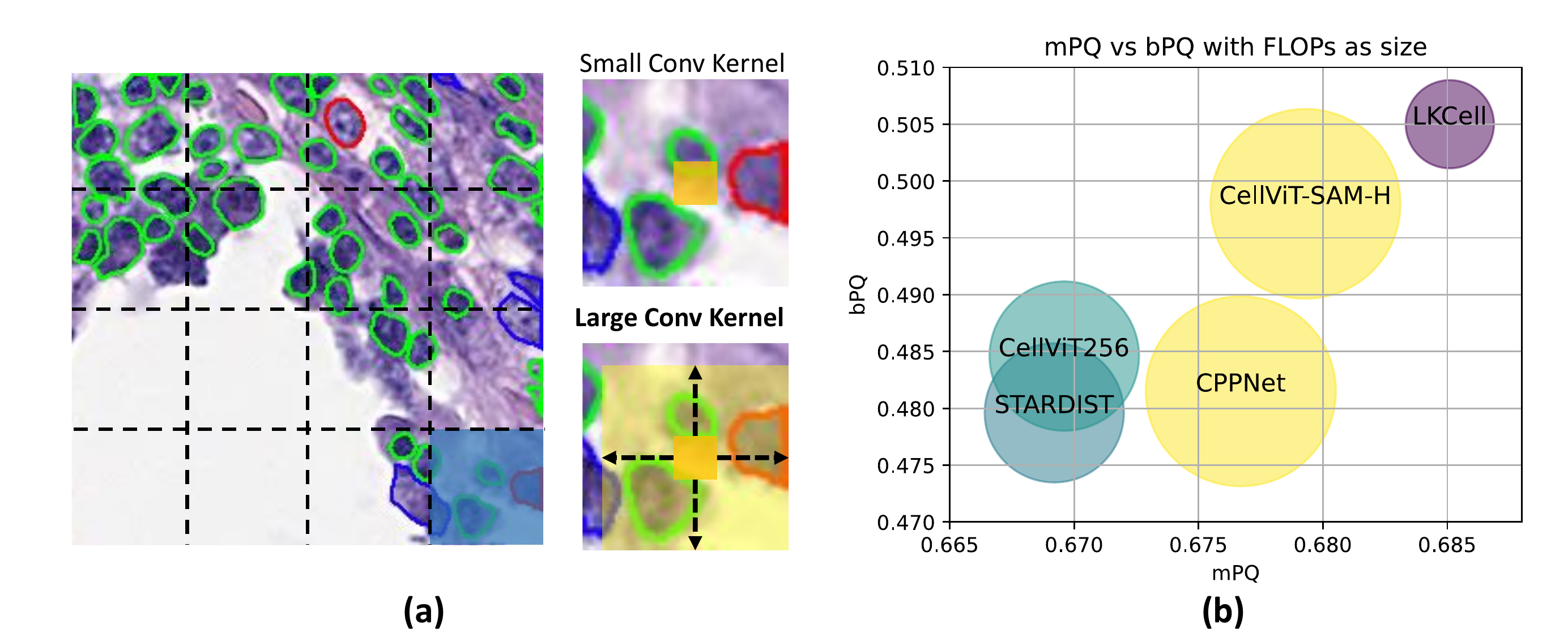}
    \caption{(a)~\textbf{Receptive Field.} By appropriately enlarging the size of the convolutional kernel, the network can effectively capture the overall structure of the cells without introducing excessive computational load. (b)~\textbf{Performance of LKCell.} We illustrate the computational efficiency and performance metrics of LKCell compared to previous methods. LKCell achieves \textit{state-of-the-art} performance with minimal FLOPs.}
    \label{fig1}
\end{figure}

In this paper, we rethink the receptive field in cell segmentation and ask: is a global receptive field with a high computational cost necessary for effective cell segmentation? As shown in Figure~\ref{fig1} (a), in tissue images stained with the blood dyes hematoxylin and eosin (H$\&$E), there are usually a certain number of cell nuclei within the field of view. We believe a receptive field capable of capturing the entire cell is crucial for successful cell segmentation, which is larger than traditional convolution but smaller than ViT. Therefore, unlike the two methods mentioned above, a new approach has emerged: achieving both high efficiency and high performance in cell segmentation may be possible by appropriately expanding the receptive field of a single convolutional kernel, i.e. large convolution kernels. The recent researches~\cite{ding2022scaling,ding2023unireplknet} introduce a similar routine to natural image analysis and achieves great success.

\begin{figure}[t]
  \centering
  \includegraphics[width=\linewidth]{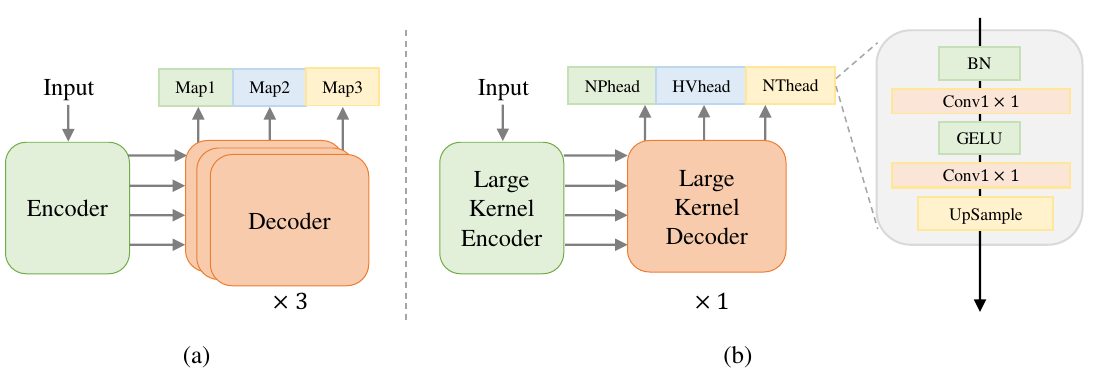}
  \caption{
  \textbf{Comparison with previous best methods.} (a) represents a typical Hover-Net~\cite{graham2019hover} shaped model consisting of three decoder branches, each producing three different maps. On the other hand, (b) represents our model, which consists of a single decoder and three separate segmentation heads for different outputs. This significantly reduces the parameter and computational complexity of the model.}
  \label{fig2:brief_lknet}
\end{figure}

Inspired by these, we propose LKCell, a method for cell nucleus segmentation based on large convolution kernels, which offers a large receptive field and efficient computation. To the best of our knowledge, we are the first to introduce a large receptive field into the field of cell nucleus segmentation. Firstly, for the feature extractor, we transfer the backbones with large kernels pre-trained on natural images~\cite{ding2023unireplknet} to the medical segmentation field.
Secondly, the previous models commonly employed three-layer decoders to obtain the output maps~\cite{graham2019hover, horst2024cellvit, tommasino2023hover}. We believe this design introduces parameter redundancy and is not necessary.
Instead, we design a single-layer decoder with large convolution kernels and connect different segmentation heads to obtain corresponding output maps. In detail, in the LKCell module of the convolution kernel, we employ multiple large convolution kernels of different sizes inspired by ~\cite{ding2022scaling}. These kernels enable the network to capture multi-scale contextual information and effectively handle significant size variations between cell nuclei and the background. We parallelly incorporate small convolution kernels into the large convolution kernels, allowing the aggregation of contextual information within the receptive field and gradually increasing the effective receptive field to extract finer and more informative features.

By unleashing the potential of large convolution kernels, our approach demonstrates significant advantages in cell nucleus instance segmentation. As illustrated in Figure~\ref{fig1} (b), compared to previous state-of-the-art methods, our method achieves a \textit{78.4$\%$} reduction in FLOPs in terms of computation efficiency while reaching the current \textit{state-of-the-art} of performance. The main contributions of the proposed method are as follows:
\begin{itemize}
\item We propose LKCell, a segmentation method based on large convolution kernels. It introduces large convolution kernels for the first time in the field of nucleus segmentation, achieving efficient and accurate nucleus segmentation. 
\item We design a novel decoder based on large convolution kernels and simplify the previous model's multi-layered design. Our method has achieved remarkable improvements in performance while successfully reducing the number of model parameters.
\item LKCell achieves \textit{state-of-the-art} results on the PanNuke dataset, with an mPQ score of \textit{0.5080} and a bPQ score of \textit{0.6847}. Extensive experiments validate the effectiveness of our proposed methodology.

\end{itemize}

%% file: related.tex
\paragraph{\textbf{Nuclei Segmentation based on CNN}}
Traditional image processing techniques~\cite{ali2012integrated, graham2019hover, koonce2021resnet} design and extract features specific to the cell segmentation. For instance, Ali el. al.~\cite{ali2012integrated} utilize predefined nuclear geometry and the watershed algorithm to separate clustered nuclei. However, these traditional techniques heavily rely on manually annotated features derived from expert-level domain knowledge, which inherently limits their representational capacity, particularly when working with scarce datasets.

Deep learning has recently emerged as the primary approach for cell nucleus segmentation. Typical Convolution Neural Networks (CNN) have~\cite{ali2012integrated, graham2019hover, koonce2021resnet, ronneberger2015u} been widely applied in medical image segmentation. Among them, the U-Net architecture~\cite{ronneberger2015u} has propelled the development of digital pathology. 
Based on them, several methods~\cite{raza2019micro,naylor2018segmentation, graham2019hover, he2016deep, chen2023cpp, schmidt2018cell} combine CNN architectures with post-processing operations to achieve automatic cell segmentation. Micro-Net~\cite{raza2019micro} involves using multi-resolution training and connects intermediate layers to improve localization and contextual information. DIST~\cite{naylor2018segmentation} formulates the segmentation problem as a regression task of distance maps to address the segmentation of overlapping nuclei. HoVerNet~\cite{graham2019hover} utilizes ResNet50~\cite{he2016deep} as the encoder and employs three decoders to obtain maps for HV, NP, and NC, followed by post-processing using the watershed algorithm to obtain instance segmentation maps. Note that it is currently one of the most widely used post-processing methods in research. CPP-Net~\cite{chen2023cpp}incorporates parallel convolution layers in the post-processing stage to predict inter-nuclear distances. STARDIST~\cite{schmidt2018cell} adopts the U-Net architecture and introduces a new approach to better match and identify star-shaped structures.

Despite the proven effectiveness of traditional CNN models in image processing, they are limited to local capabilities and may struggle to capture long-range spatial relationships~\cite{ester2023valuing}.
Constrained by the receptive field, these methods have limited performance. In this paper, we mainly explore the impact of enlarging the convolution kernel on cell segmentation.


\paragraph{\textbf{Nuclei Segmentation based on based on ViT}}
Few works have introduced Transformers~\cite{vaswani2017attention} into nuclei segmentation to improve the models' ability to capture global information.
Trans-Unet~\cite{chen2021transunet} is a model that combines the advantages of Transformers and U-Net. It uses Transformers to encode segmented image patches from CNN feature maps into input sequences, extracting global context information. 
Swin-Unet~\cite{cao2022swin} is a model that utilizes Swin Transformer~\cite{liu2021swin} blocks and adopts a symmetric encoder-decoder structure with skip connections. 
A recent state-of-the-art method, CellViT~\cite{horst2024cellvit}, follows the approach of UNETR~\cite{hatamizadeh2022unetr} and utilizes Vision Transformers (ViT)~\cite{dosovitskiy2021image} as the backbone at the 2D level. This work marks the first introduction of ViT into the field of cell nucleus segmentation. 
However, the global receptive field of ViT also introduces a significant computational burden. This makes it challenging to be widely used in clinical applications. In this paper, we rethink the relationship and necessity between the receptive field and cell segmentation.

\paragraph{\textbf{Architecture based on Large concolution kernels}}
In natural images, a different approach has emerged.
Some researchers~\cite {liu2022convnet,woo2023convnext} have proposed networks based on large convolution kernels.~\cite{azad2024beyond} applies a U-net-shaped network structure and utilizes large convolution kernels to obtain a broader receptive field.RepLKNet~\cite{ding2022scaling}has achieved very excellent results in the semantic segmentation task of natural scenes.UniRepLKNet~\cite{ding2023unireplknet} has demonstrated outstanding performance across various modalities.
In this paper, we introduce this approach to the field of cell segmentation for the first time. We find that large convolution kernels demonstrate remarkable potential in cell segmentation.

%% file: methods.tex
In this section, we provide an overview of our approach. Firstly, we review the underlying concept of large convolution kernels. Subsequently, we introduce our innovative ideas and exploratory efforts concerning the utilization of these large convolution kernels. Building upon these advancements, we unveil the network architecture tailored specifically for the task of cell nucleus instance segmentation.
\begin{figure}[t]
  \centering
  \includegraphics[width=\linewidth]{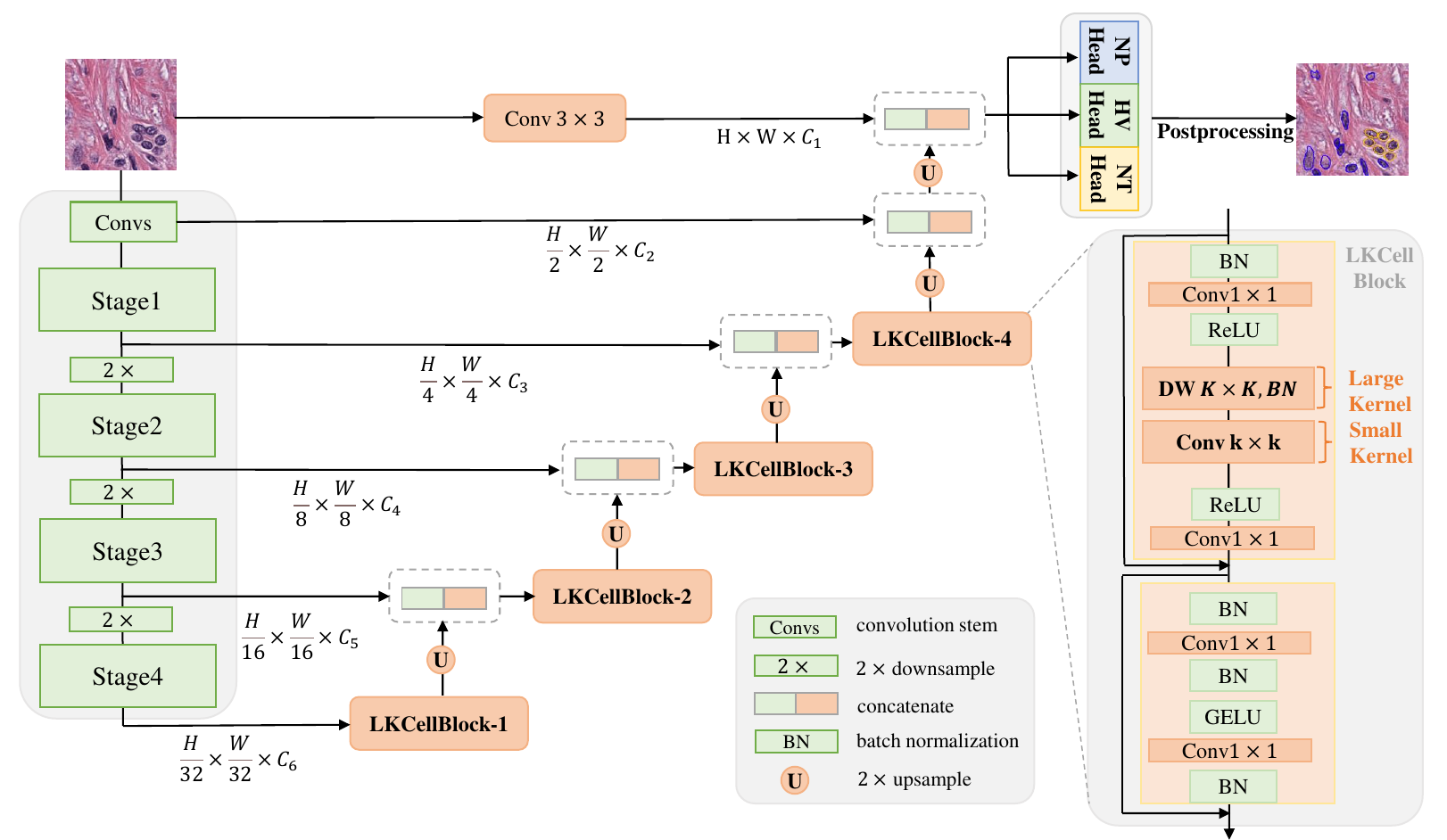}
  \caption{
  \textbf{Architeture of LKCell.} We present the overall architecture of LKCell. The encoder is composed of a pre-trained model~\cite{ding2023unireplknet} with large convolution kernels and is connected to the decoder through skip connections. The decoder consists of four LKCellBlocks. Each LKCellBlock is a combination of Large Kernel and Small Kernel, along with components such as BatchNorm, GELU, ReLU, and 1$\times$1 convolution. Postprocessing technique is employed to match nuclei types and refine nuclei segments.
  }
  \label{fig1:sample}
\end{figure}

\subsection{Large convolution Kernel}

For ease of understanding, we first introduce the basic design principles of large convolution kernels.

Large convolution kernels provide a sufficiently large receptive field and effectively aggregate spatial information, aiding in learning the relative positions between concepts and encoding absolute positional information through the padding effect. Structural reparameterization \cite{ding2021repvgg} refers to replacing a large convolution kernel with multiple small convolution kernels to reduce the number of parameters and computations. To construct large convolution kernels, techniques such as Depth-Wise Convolution (DW Conv)~\cite{howard2017mobilenets}, parallel $K \times K$ depth-wise convolution for structural reparameterization, and adding $1 \times 1 $ convolution before depth-wise convolution to increase feature dimensionality and enhance non-linearity and inter-channel information exchange can be employed, effectively reducing the number of parameters and computations. The kernel size of a depth-wise convolution can be expressed by the formula:
\begin{equation}
K_D = (2d - 1) \times (2d - 1) 
\end{equation}
where $d$ is the dilation rate.

The kernel size of a depth-wise convolution paralleled with a small depth-wise convolution and subjected to structural reparameterization can be expressed by the formula:
\begin{equation}
    K_{Drep} = (2d - 1) \times (2d - 1) \times m \times k \times k
\end{equation}
where $d$ is the dilation rate of the depth-wise convolution, $m$ is the number of convolution kernels used for structural reparameterization, and $k$ is the size of the convolution kernels used for structural reparameterization.

\subsection{Cell Segmentation with Large Convolution Kernels}
We propose LKCell, a novel architecture that integrates large convolution kernels~\cite{ding2023unireplknet}. It leverages the advantages of large kernels for image encoding while preserving fine-grained information. To the best of our knowledge, this is the first time that large kernel networks have been introduced to cell nucleus segmentation tasks, leveraging the large convolution kernels pre-trained model~\cite{ding2023unireplknet}. Our model overview is shown in the Figure \ref{fig1:sample}.

\subsubsection{LKCell Block}

We design a Large Kernel Cell Block (LKCell Block). It is the basic module of our model. As shown in Figure~\ref{fig1:sample}, this block leverages small kernels and multiple dilated small kernel layers to enhance the capturing ability of non-dilated large kernel convolution layers, thereby obtaining higher-quality features. Specifically, the block consists of a large kernel convolution layer with a kernel size of $K$ and $n$ parallel convolution layers with a dilation rate of $r$, satisfying $(n-1)r+1 \leq K$. This design enables the simultaneous capture of small-scale and large-scale patterns.

\subsubsection{Transfer Pretrained LK-Encoder to Cell Segmentation}
\label{sec:encoder}
Our encoder mainly follows the success design of~\cite{ding2023unireplknet} consisting of four stages.
We experiment with two model variants, namely LKCell-B and LKCell-L. The corresponding parameter config can be found in Table~\ref{tab:model}.
Differently, inspired by U-Net architecture, our encoder has five output branches, including four stage outputs and an additional output from the first downsampling block. This design enables the encoder to fully utilize the depth information and provide more low-level features.

\begin{table}
\footnotesize
  \caption{\textbf{Network Configurations.}}\label{tabar}%
  \label{tab:model}
  \centering
  \begin{tabular}{l|ccccccc}
    \toprule
    \bf Model & \bf N1 & \bf N2 & \bf N3 & \bf N4 & \bf C&\bf Params(M)&\bf FLOPs(G) \\
    \midrule
    LKCell-B & 2 & 2 &  8 & 2 & (64, 128, 256, 512)&122.53&46.25 \\
    LKCell-L & 3 & 3 & 27 & 3 & (64, 128, 256, 512)&163.84&47.86 \\
    \bottomrule
  \end{tabular}
\end{table}

\subsubsection{You only need one LK-Decoder}

We have made two contributions to the cell segmentation decoder design. First, we clarify and simplify the redundancy issues present in the design of previous methods~\cite{horst2024cellvit, graham2019hover}. Second, we design the first cell segmentation decoder based on large convolution kernels to enhance network performance.

Our network incorporates a single decoder and three distinct multitask output branches for segmentation maps, drawing inspiration from HoVer-Net~\cite{graham2019hover}. However, we believe that the three identical decoder branches in HoVer-Net introduce parameter redundancy. To tackle this issue, we propose a decoder that utilizes large convolution kernels and upsampling. Following the U-Net architecture, our design maintains symmetry by consisting of four stages that correspond to the four stages of the encoder. Due to the loss of spatial information caused by downsampling in the encoder, our training approach incorporates multi-resolution input images and connects intermediate layers to improve localization and contextual information. Simultaneously, it adapts the U-Net network to accommodate nuclei of different sizes in the output. This fusion of features aims to minimize the spatial information loss resulting from downsampling in the encoder.

Within a single decoder stage, we apply the LKCell block to the decoder features from the previous stage to introduce non-linearity and promote information exchange across channels, while reducing computation and parameter count. We then perform upsampling and match the upsampled features with the corresponding skip-connection features. By concatenating these upsampled features with the skip-connection features, we obtain semantically and spatially rich features. For our model, in the \textit{i-th} decoder stage, let $F^{i-1}$ represent the features from the previous decoder stage with dimensions $ c^{i-1} \times h \times w $. Similarly, let $Z^{i}$ represent the skip-connection features from the same stage with dimensions $ c^{i} \times 2h \times 2w $. We can express the operations for each decoder stage using the following equation:
\begin{equation}
    F^{i} = LKCellBlock(F^{i-1})
\end{equation}
\begin{equation}
    F^{i} = UPCat(F^{i},Z^{i})
\end{equation}

After the final decoder stage, we establish a direct connection between the input image and the decoder output using convolution layers to create a skip connection. This skip connection is then fused with the output features from the last stage to generate the final three segmentation maps. This skip connection enables a direct flow of information from the input image to the segmentation output, enhancing the overall segmentation performance.

\subsection{Postprocessing}
Since the network itself cannot directly provide instance-level segmentation of individual cell nuclei, postprocessing is required to obtain accurate results. 
The postprocessing mainly involves merging results from different segmentation maps, separating overlapping nuclei to ensure more precise individual nucleus segmentation, and determining the types of nuclei based on our nucleus type map.
The nuclei class is determined using a postprocessing method inspired by HoVer-Net. As the boundaries between nuclei and background exhibit significant gradient changes, we compute the gradients of the horizontal and vertical distance maps to capture the transformations at the nucleus boundaries and background edges. 
The Sobel operator is then employed to identify regions with significant changes. 
This allows for the separation of adjacent nuclei and overlapping nuclei. Finally, marker-controlled watershed \cite{castleman1996digital} is applied to generate the final boundaries of the cell nuclei. The nucleus type map is utilized to perform majority voting within the nuclei regions, assigning the majority class to the separated nuclei images. This method aims to improve the accuracy and consistency of cell nucleus type predictions.

%% file: exp.tex
\begin{sidewaystable}
\small
\centering
\caption{\textbf{Average mPQ and bPQ}. Average mPQ and bPQ values are obtained by each model on the PanNuke dataset using three-fold cross-validation for the 19 tissue types. The overall average mPQ and bPQ values for the 19 tissue types are also provided. Please note that TSFD-Net is not evaluated on the official three-fold splits of the PanNuke dataset and is excluded from the comparison. The experimental results demonstrate that our model not only achieves the best performance in terms of overall mPQ and bPQ, but also maintains excellent performance across all 19 nuclei classes. This highlights the robustness of our model.}\label{tab3}
\vspace{0.5em}
\scalebox{0.9}{
 \begin{tabular}
{l@{\hspace{0.15cm}}c@{\hspace{0.18cm}}c@{\hspace{0.18cm}}c@{\hspace{0.18cm}}c@{\hspace{0.18cm}}c@{\hspace{0.07cm}}c@{\hspace{0.18cm}}c@{\hspace{0.18cm}}c@{\hspace{0.18cm}}c}

    \toprule                  
    Model  &  HoVer-Net\cite{graham2019hover} & STARDIST\cite{schmidt2018cell} & TSFD-Net\cite{ilyas2022tsfd} &  CPP-Net\cite{chen2023cpp} & CellViT-256\cite{horst2024cellvit} & CellViT-SAM-H\cite{horst2024cellvit} & \bfseries Ours-B &  \bfseries 
    & \bfseries Ours-L \\
    \midrule
    Tissue & mPQ\hspace{0.09cm} bPQ & mPQ\hspace{0.09cm} bPQ & mPQ \hspace{0.09cm}  bPQ & mPQ\hspace{0.09cm} bPQ & mPQ\hspace{0.09cm} bPQ &  mPQ\hspace{0.09cm} bPQ & mPQ\hspace{0.09cm} bPQ & 
    & mPQ\hspace{0.09cm} bPQ     \\[1ex]
    \hline
    Adrenal &0.4812\hspace{0.09cm} 0.6962 & 0.4868\hspace{0.09cm} 0.6972 & 0.5223 \hspace{0.09cm} 0.6900 & 0.4922\hspace{0.09cm} 0.7031 & 0.4950\hspace{0.09cm} 0.7009& \quad \underline{0.5134}\hspace{0.09cm} 0.7086 & 0.5032\hspace{0.09cm} \bf0.7203 & 
    & \underline{0.5077}\hspace{0.09cm} \underline{0.7150} \\[1ex]
    
    Bile\_Duct & 0.4714\hspace{0.1cm} 0.6696 & 0.4651\hspace{0.1cm} 0.6690 & 0.5000 \hspace{0.1cm} 0.6284 & 0.4650\hspace{0.1cm} 0.6739 & 0.4721\hspace{0.1cm} 0.6705 & \quad \underline{0.4887}\hspace{0.1cm} 0.6784 & 0.4817\hspace{0.1cm} \underline{0.6811} & 
    & \bf0.5155\hspace{0.1cm} 0.6961 \\[1ex]
    
    Bladder & 0.5792\hspace{0.1cm} 0.7031 & 0.5793\hspace{0.1cm} 0.6986 & 0.5738 \hspace{0.1cm} 0.6773 & 0.5932\hspace{0.1cm} 0.7057 & 0.5756\hspace{0.1cm} 0.7056 & \quad 0.5844\hspace{0.1cm} 0.7068 & \bf0.6056\hspace{0.1cm} 0.7155 & 
    & \underline{0.6011}\hspace{0.1cm} \underline{0.7141} \\[1ex]
    
    Breast  & 0.4902\hspace{0.1cm} 0.6470 & 0.5064\hspace{0.1cm} 0.6666 & 0.5106 \hspace{0.1cm} 0.6245 & 0.5066\hspace{0.1cm} 0.6718 & 0.5089\hspace{0.1cm} 0.6641 & \quad \underline{0.5180}\hspace{0.1cm} \textbf{0.6748} & \textbf{0.5194}\hspace{0.1cm} 0.6701 
    &
    & 0.5143\hspace{0.1cm} \underline{0.6723} \\[1ex]
    
    Cervix & 0.4438\hspace{0.1cm} 0.6652 & 0.4628\hspace{0.1cm} 0.6690 & 0.5204 \hspace{0.1cm} 0.6561 & 0.4779\hspace{0.1cm} 0.6880 & 0.4893\hspace{0.1cm} 0.6862 & \quad 0.4984\hspace{0.1cm} 0.6872 & \textbf{0.5114}\hspace{0.1cm} \textbf{0.6993} & 
    & \underline{0.5021}\hspace{0.1cm} \underline{0.6951} \\[1ex]
    
    Colon & 0.4095\hspace{0.1cm} 0.5575 & 0.4205\hspace{0.1cm} 0.5779 & 0.4382 \hspace{0.1cm} 0.5370 & 0.4269\hspace{0.1cm} 0.5888 & 0.4245\hspace{0.1cm} 0.5700 & \quad 0.4485\hspace{0.1cm} \underline{0.5921} & \underline{0.4496}\hspace{0.1cm} 0.5905 & 
    & \bf0.4637\hspace{0.1cm} \bf0.6013 \\[1ex]
    
    Esophagus & 0.5085\hspace{0.1cm} 0.6427 & 0.5331\hspace{0.1cm} 0.6655 & 0.5438 \hspace{0.1cm} 0.6306 & 0.5410\hspace{0.1cm} \underline{0.6755} & 0.5373\hspace{0.1cm} 0.6619 & \quad 0.5454\hspace{0.1cm} 0.6682 & \underline{0.5577}\hspace{0.1cm} \bf0.6821 & 
    & \textbf{0.5593}\hspace{0.1cm} 0.6744 \\[1ex]
    
    Head\&Neck & 0.4530\hspace{0.1cm} 0.6331 & 0.4768\hspace{0.1cm} 0.6433 & 0.4937 \hspace{0.1cm} 0.6277 & 0.4667\hspace{0.1cm} 0.6468 & 0.4901\hspace{0.1cm} 0.6472 & \quad 0.4913\hspace{0.1cm} 0.6544 & \underline{0.5068}\hspace{0.1cm} \underline{0.6624} & 
    & \bf0.5278\hspace{0.1cm} \bf0.6715\\[1ex]
    
    Kidney & 0.4424\hspace{0.1cm} 0.6836 & \textbf{0.5880}\hspace{0.1cm} 0.6998 & 0.5517 \hspace{0.1cm} 0.6824 & 0.5092\hspace{0.1cm} 0.7001 & 0.5409\hspace{0.1cm} 0.6993 & \quad 0.5366\hspace{0.1cm} 0.7092 &0.5516\hspace{0.1cm} \underline{0.7168} & 
    & \underline{0.5735}\hspace{0.1cm} \bf0.7275 \\[1ex]
    
    Liver & 0.4974\hspace{0.1cm} 0.7248 & 0.5145\hspace{0.1cm} 0.7231 & 0.5079 \hspace{0.1cm} 0.6675 & 0.5099\hspace{0.1cm} 0.7271 & 0.5065\hspace{0.1cm} 0.7160 & \quad 0.5224\hspace{0.1cm} 0.7322 & \underline{0.5282}\hspace{0.1cm} \underline{0.7387} & 
    & \bf0.5350\hspace{0.1cm} \bf0.7428 \\[1ex]
    
    Lung  & 0.4004\hspace{0.1cm} 0.6302 & 0.4128\hspace{0.1cm} 0.6362& 0.4274 \hspace{0.1cm} 0.5941 & 0.4234\hspace{0.1cm} 0.6364 & 0.4102\hspace{0.1cm} 0.6317 & \quad \underline{0.4314}\hspace{0.1cm} 0.6426 & 0.4307\hspace{0.1cm} \bf0.6538 & 
    & \textbf{0.4415}\hspace{0.1cm} \underline{0.6458} \\[1ex]
    
    Ovarian  & 0.4863\hspace{0.1cm} 0.6309 & 0.5205\hspace{0.1cm} 0.6668 & 0.5253 \hspace{0.1cm} 0.6431 & 0.5276\hspace{0.1cm} \underline{0.6792} & 0.5260\hspace{0.1cm} 0.6596 & \quad \underline{0.5390}\hspace{0.1cm} 0.6722 & \bf0.5471\hspace{0.1cm} \bf0.6815 & 
    & 0.5311\hspace{0.1cm} 0.6672 \\[1ex]
    
    Pancreatic  & 0.4600\hspace{0.1cm} 0.6491 & 0.4585\hspace{0.1cm} 0.6601 & 0.4893 \hspace{0.1cm} 0.6241 & 0.4680\hspace{0.1cm} 0.6742 & 0.4769\hspace{0.1cm} 0.6643 & \quad 0.4719\hspace{0.1cm} 0.6658 & \bf0.5133\hspace{0.1cm} \bf0.6769 & 
    & \underline{0.4795}\hspace{0.1cm} \underline{0.6730} \\[1ex]
    
    Prostate  & 0.5101\hspace{0.1cm} 0.6615 & 0.5067\hspace{0.1cm} 0.6748 & 0.5431 \hspace{0.1cm} 0.6406 & 0.5261\hspace{0.1cm} \textbf{0.6903} & 0.5164\hspace{0.1cm} 0.6695 & \quad \textbf{0.5321}\hspace{0.1cm} 0.6821 & 0.5266\hspace{0.1cm} \underline{0.6870} & 
    & \underline{0.5316}\hspace{0.1cm} 0.6781 \\[1ex]
    
    Skin  & 0.3429\hspace{0.1cm} 0.6234 & 0.3610\hspace{0.1cm} 0.6289 & 0.4354 \hspace{0.1cm} 0.6074 & 0.3547\hspace{0.1cm} 0.6192 & 0.3661\hspace{0.1cm} 0.6400 & \quad \bf0.4339\hspace{0.1cm} \underline{0.6565} & 0.4183\hspace{0.1cm} 0.6437 & 
    & \underline{0.4217}\hspace{0.1cm} \bf0.6662 \\[1ex]
    
    Stomach  & \textbf{0.4726}\hspace{0.1cm} 0.6886 & 0.4477\hspace{0.1cm} 0.6944 & 0.4871 \hspace{0.1cm} 0.6529 & 0.4553\hspace{0.1cm} 0.7043 & 0.4475\hspace{0.1cm} 0.6918 & \quad \underline{0.4705}\hspace{0.1cm} 0.7022 &0.4620\hspace{0.1cm} \bf0.7101 & 
    & 0.4506\hspace{0.1cm} \underline{0.7057} \\[1ex]
    
    Testis  & 0.4754\hspace{0.1cm} 0.6890 & 0.4942\hspace{0.1cm} 0.6869 & 0.4843 \hspace{0.1cm} 0.6435 & 0.4917\hspace{0.1cm} \underline{0.7006} & 0.5091\hspace{0.1cm} 0.6883 & \quad \underline{0.5127}\hspace{0.1cm} 0.6955 & \bf0.5273\hspace{0.1cm} \bf 0.7101 & 
    & 0.5091\hspace{0.1cm} 0.6979\\[1ex]
    
    Thyroid  & 0.4315\hspace{0.1cm} 0.6983 & 0.4300\hspace{0.1cm} 0.6962 & 0.5154 \hspace{0.1cm} 0.6692 & 0.4344\hspace{0.1cm} 0.7094 & 0.4412\hspace{0.1cm} 0.7035 & \quad 0.4519\hspace{0.1cm} \underline{0.7151} & 0.4673\hspace{0.1cm}\bf0.7161 & 
    & \textbf{0.4698}\hspace{0.1cm} 0.7037\\[1ex]
    
    Uterus  & 0.4393\hspace{0.1cm} 0.6393 & 0.4480\hspace{0.1cm} 0.6599 & 0.5068 \hspace{0.1cm} 0.6204 & 0.4790\hspace{0.1cm} \underline{0.6622} & 0.4737\hspace{0.1cm} 0.6516 & \quad 0.4737\hspace{0.1cm} 0.6625 & \underline{0.4879}\hspace{0.1cm} \textbf{0.6681} & 
    & \textbf{0.5167}\hspace{0.1cm} 0.6616 \\[2ex]
    
    \hline
    \bfseries Average  & 0.4629\hspace{0.1cm} 0.6596 & 0.4796\hspace{0.1cm} 0.6692 & 0.5040\hspace{0.1cm} 0.6377 & 0.4815\hspace{0.1cm} 0.6767 & 0.4846\hspace{0.1cm} 0.6696 & \quad 0.4980\hspace{0.1cm} 0.6793 & \underline{0.5050}\hspace{0.1cm} \bf 0.6851 & 
    & \textbf{0.5080}\hspace{0.1cm} \underline{0.6847} \\
\hline
\end{tabular}}
\end{sidewaystable}

\subsection{Datasets}
\paragraph{PanNuke}
The PanNuke dataset~\cite{gamper2019pannuke} comprises H$\&$E stained images with a resolution of 256×256 pixels, totaling 7,904 images from 19 different tissue types. Within these images, cell nuclei are classified into five distinct cell categories: neoplastic cells, inflammatory cells, connective cells, dead cells, and epithelial cells. Due to the imbalanced distribution of cell types, the PanNuke dataset is considered one of the most challenging datasets for cell nucleus instance segmentation tasks. To address this issue, our model follows the training and evaluation guidelines outlined in ~\cite{gamper2020pannuke} and employs a three-fold cross-validation approach. The dataset is divided into three folds, with one fold used for training the model and the remaining two folds used for evaluation and inference. This division helps facilitate effective model training and enables robust evaluation across different dataset partitions.

\paragraph{MoNuSeg}

The MoNuSeg dataset~\cite{8880654} consists of H\&E stained tissue images captured at a 40$\times$ magnification. It includes a training set of 30 images and a test set of 14 images. The images have a size of 1000$\times$1000 and are sampled from different whole-slide slices of various organs. Compared to the PanNuke dataset, MoNuSeg is much smaller and does not have fine-grained classes for cell nuclei. Therefore, in our experiments, we only use this dataset as the test dataset.

\subsection{Metric}
Cell nuclei instance segmentation not only requires accurate recognition of each nucleus's location but also necessitates distinguishing individual nuclei. Therefore, the evaluation metrics need to simultaneously satisfy both separating nuclei from the background and detecting individual nuclei instances and segmenting each instance. We adopt Panoptic Quality (PQ) as the evaluation metric, as suggested by the PanNuke dataset evaluation protocol. PQ takes into account not only the accuracy of instance detection and classification but also the quality of instance segmentation, providing a more comprehensive quantitative metric.

Panoptic Quality (PQ) is an intuitive and comprehensive metric that can be decomposed into two components: Detection Quality (DQ) and Segmentation Quality (SQ). The DQ evaluates the model's accuracy in recognizing and localizing instances, similar to the $F_1$ score in classification and detection scenarios. The SQ assesses the model's performance in accurately segmenting instance boundaries. The PQ metric is calculated as the product of DQ and SQ, providing a more comprehensive evaluation of instance segmentation performance. Mathematically:

\begin{equation}
    PQ = DQ \times SQ
\end{equation}
The Detection Quality (DQ) metric evaluates the model's detection performance. It is calculated as the ratio of the number of true positive instances (TP) to the sum of true positive instances, half of the number of false positive instances (FP), and half of the number of false negative instances (FN).

\begin{equation}
    DQ = TP/(TP +0.5FP + 0.5FN)
\end{equation}

The Segmentation Quality (SQ) metric evaluates the model's segmentation performance. It is calculated as the average Intersection over Union (IoU) of all detected instances, i.e., the sum of IoUs of all correctly detected instances divided by the number of true positive instances (TP).
\begin{equation}
    SQ= (\sum IoU(y,\hat{y}))/{TP}
\end{equation}
where $IoU(y,\hat{y})$ denotes the Intersection over Union. Here, $y$ represents the ground truth of the correctly segmented instance, $\hat{y}$ represents the predicted segmentation, and $(y,\hat{y})$ represents the intersection of the correctly segmented instance and the predicted segmentation.

Considering the diversity of classes in this dataset, we adopt two types of Panoptic Quality (PQ) scores: Binary Panoptic Quality (bPQ): separates cell nuclei from the background, analogous to a traditional binary classification problem. Multi-class Panoptic Quality (mPQ): independently calculates the PQ score for each class of cell nuclei, and averages the results across all classes. To evaluate the model's performance in detecting cell nuclei (i.e., to assess the effectiveness of our model on the MoNuSeg dataset), we also employ conventional detection metrics. The evaluation metrics include:
\begin{equation}
    Dice = {2TP}/{(2TP + FP + FN)}
\end{equation}
\begin{equation}
    F_1={2TP_m}/{(2TP_m + FP_m + FN_m)}
\end{equation}
The evaluation metrics include:
$Dice$the Dice coefficient is used to measure the similarity between predicted and true segmentation results. It ranges from 0 to 1, where a value closer to 1 indicates a higher degree of overlap between the segmentation results and the ground truth.
$F_1$ score: the harmonic mean of precision and recall, providing a balanced measure of detection performance.
True Positives ($TP_m$): correctly detected instances.

False Positives ($FP_m$): instances misclassified as positive, indicating errors in detection.

False Negatives ($FN_m$): undetected instances, highlighting missed opportunities for detection.

\subsection{Results}
This section presents a comprehensive evaluation of our approach, highlighting the segmentation quality of the PanNuke dataset, as well as the generalization capabilities of the MoNuSeg dataset.

\begin{figure}[t]
  \centering
  \includegraphics[width=1\linewidth]{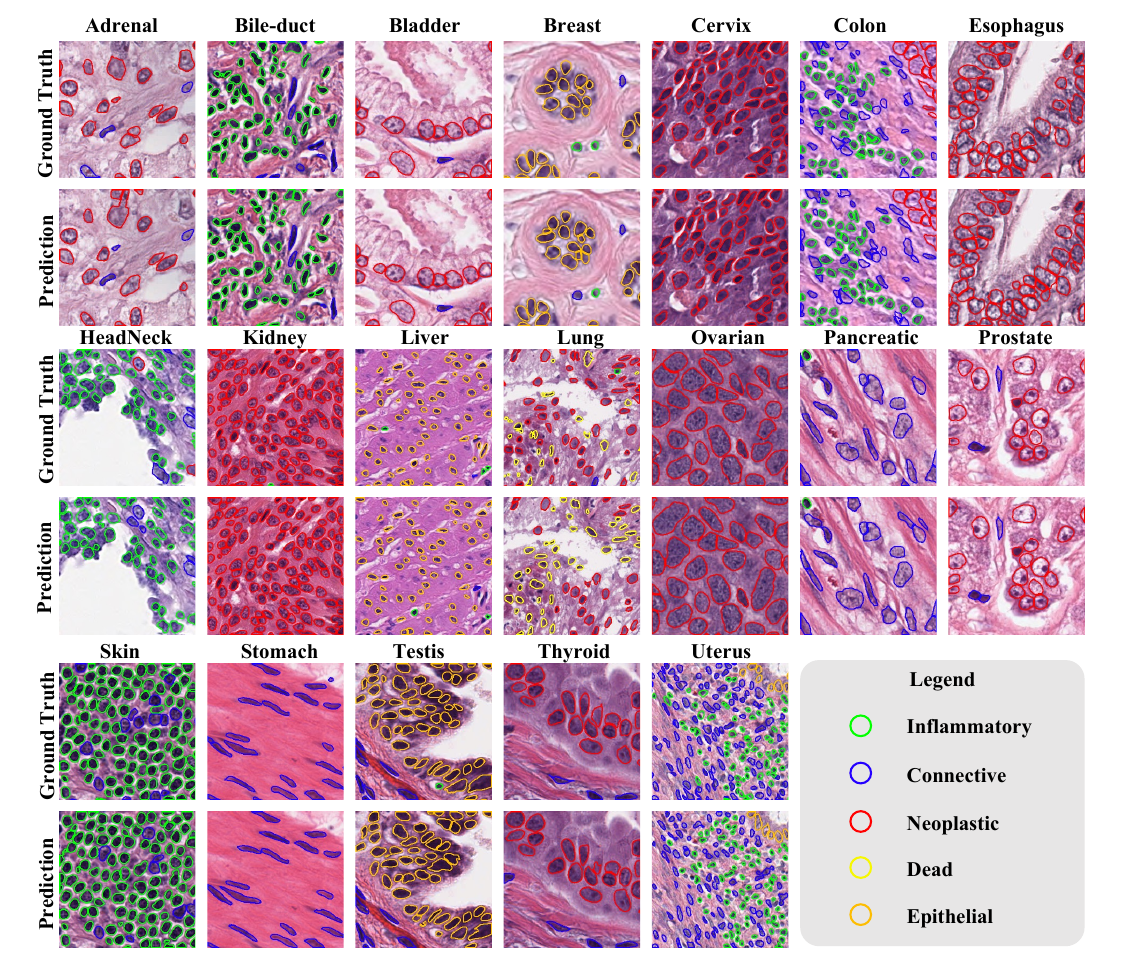}
  \caption{
  \textbf{Comparison of Segmentation Results.}We compare the segmentation results of 19 different types of cell nuclei using LKCell on the PanNuke dataset with Ground Truth and obtain highly accurate instance segmentation results.}
  \label{fig:performence}
\end{figure}

\begin{table}
\footnotesize
  \caption{\textbf{Performence on PanNuke}. Average Panoptic Quality (PQ) values for each nuclei class in the PanNuke dataset using three-fold cross-validation. The experimental results indicate that our model achieves optimal performance in terms of Panoptic Quality (PQ) for each nucleus class, while consuming only 20$\%$ of the computational load of CellViT-SAM-H~\cite{horst2024cellvit}.}
  \label{sample-table}
  \centering
  \scalebox{0.72}{
  \renewcommand{\arraystretch}{1.0} 
  \begin{tabular}{lccccccc}
    \toprule                 
    \textbf{Method} &\textbf{Params(M)}&\textbf{FLOPs(G)}  &\textbf{Neoplastic} & \textbf{Inflammatory}    &\textbf{Dead}  & \textbf{Connective} & \textbf{Epithelial} \\
    \hline
    DIST\cite{naylor2018segmentation} & -&-&0.439     &0.343 &0.000 &0.275 & 0.290 \\
    Mask-RCNN\cite{he2017mask} &  -&-&0.472    & 0.290 & 0.069&0.300 &0.403 \\
    Micro-Net\cite{raza2019micro} &  -&-&0.504  &0.333 &0.051 &0.334 &0.442 \\
    HoVer-Net\cite{graham2019hover} &  -&-&0.551  &0.417 &0.139 &0.388 &0.491 \\
    CellViT$_{256}$\cite{horst2024cellvit} &\bf46.75&132.89&0.567&0.405&0.144&0.405&0.559 \\
    CellViT-SAM-H\cite{horst2024cellvit} & 699.74&214.33&0.581&0.417&0.149&\bf0.423&\underline{0.583}\\
    \midrule
    \bf LKCell-B (Ours) &\underline{122.53}&\bf46.25&\underline{0.585} &\bf0.440 &\underline{0.144} &0.414&0.579 \\
    \bf LKCell-L (Ours) &163.84&\underline{47.86} &\bf0.586 &\underline{0.438} &\bf0.172 &\underline{0.417} &\bf0.584 \\
    \bottomrule
  \end{tabular}}
\end{table}

\subsubsection{Segmentation Quality of PanNuke}
To evaluate the instance segmentation quality of the model, we use the binary Panoptic Quality (bPQ) for 19 tissue types in the PanNuke dataset, which is considered a highly challenging multi-class Panoptic Quality (mPQ), and the Panoptic Quality (PQ) for each cell nucleus type. In Table \ref{tab3}, we evaluate the performance of STARDIST, CPP-Net, CellViT$_{256}$, and CellViT-SAM-H, which are models provided by HoVer-Net, TSFD-Net, and CellViT using the ResNet50 encoder. The experiments demonstrate that our model achieved superior mPQ and bPQ, showcasing excellent generalization across different tissue types.

In Table \ref{sample-table}, we present the PQ values for each cell nucleus type, which are the average values across all tissue types, providing a comprehensive evaluation of segmentation quality. 
Our model performs exceptionally well on Neoplastic, Inflammatory, Dead, Connective, and Epithelial nuclei. 
However, it can be observed that among all the models, the PQ values for dead nuclei are the lowest. This can be attributed to the class imbalance in the dataset and the small size of dead nuclei.

\subsubsection{Testing of MoNuSeg}
Given the limited size of the MoNuSeg dataset, we conduct tests on this dataset to assess the generalization capability of our model. Table \ref{tab4} presents the model's instance segmentation performance using metrics such as F1 score and Dice coefficient. Our model has achieved comparable F1 scores to the current {state-of-the-art} (SOTA) and surpassed the SOTA in terms of the Dice coefficient, showcasing its stable segmentation performance.

\begin{table}[t]
\footnotesize
\centering
\caption{\textbf{Performence on MoNuSeg}. Performance of our models of different sizes on the MoNuSeg dataset in terms of $F_1$,$Dice$. The experimental results demonstrate that our model achieved excellent performance on the $Dice$ metric, indicating that our model excels in predicting boundaries.}\label{tab4}%
\renewcommand{\arraystretch}{1.0} 
\scalebox{0.9}{
\begin{tabular}{@{}l|lll@{}}
\toprule
    \hspace{0.3cm}\bf Method &\hspace{0.3cm} $\bf F_1$ &\hspace{0.3cm}$\bf Dice$\hspace{0.3cm} \\
    \midrule
    \hspace{0.3cm}U-Net\cite{ronneberger2015u} &\hspace{0.3cm}79.43&\hspace{0.3cm}65.99\hspace{0.3cm}   \\
    \hspace{0.3cm}U-Net++\cite{zhou2018unet++} &\hspace{0.3cm}79.49&\hspace{0.3cm}66.17\hspace{0.3cm}   \\
    \hspace{0.3cm}Med Transformer \cite{wang2024medtransformer} &\hspace{0.3cm}79.55 &\hspace{0.3cm}66.17\hspace{0.3cm} \\
    \hspace{0.3cm}Swin-unet \cite{cao2022swin} &\hspace{0.3cm}79.56&\hspace{0.3cm}64.71\hspace{0.3cm}\\
    \hspace{0.3cm}MaxViT-UNet\cite{khan2023maxvit}&\hspace{0.3cm}83.78 &\hspace{0.3cm}72.08\hspace{0.3cm} \\
    \hspace{0.3cm}CellViT-SAM-H\cite{horst2024cellvit} &\hspace{0.3cm}\bf86.8 &\hspace{0.3cm}83.08\hspace{0.3cm} \\
    \hspace{0.3cm}\bf LKCell-L (Ours)&\hspace{0.3cm}82.99&\hspace{0.3cm}\bf 83.96\hspace{0.3cm} \\
    \bottomrule
  \end{tabular}
}
\end{table}

\subsubsection{Ablation Study of the Proposed LKCell}

To evaluate the effectiveness of our proposed model, we separately connect ResNet50 as the encoder to both a conventional U-Net decoder and our proposed decoder.
The experimental results demonstrate a significant improvement in instance segmentation performance with our decoder. 
Moreover, the utilization of a backbone network with larger convolution kernels outperforms the original CNN network.

Additionally, we compare the performance of using ViT as the encoder with that of using a conventional U-Net encoder or our decoder. The experiments show poor performance of ViT, and further analysis indicates that ViT's performance heavily relies on pretraining on large-scale datasets, which contrasts with the limitations of our relatively small medical dataset. 
Notably, when our decoder is connected to the ViT encoder, the mPQ increase by 0.0899 and the bPQ increase by 0.1119 compared to using the conventional U-Net decoder, highlighting the effectiveness of our proposed decoder with larger convolution kernels. See Table~\ref{tab5}.

\begin{table}[t]
\footnotesize
\caption{\textbf{Ablations of LKCell.} We ablate our decoder in different architectures~\cite{koonce2021resnet, touvron2021training, ding2023unireplknet}. The experimental results demonstrate that: (1) Our decoder can adapt to various structures, enhancing their segmentation performance. (2) With the enhancement of the decoder, the traditional multi-decoder design~\cite{horst2024cellvit, graham2019hover} is no longer necessary. The property significantly reduces the network's computational load and number of parameters. }\label{tab5}%
\vspace{0.5em}
\renewcommand{\arraystretch}{1.1} 
\scalebox{0.85}{
\begin{tabular}{@{}l|llllc@{}}
\toprule                 
    \bf Method &\bf FLOPs(G)   & \bf Params(M) & \bf mPQ $\uparrow$ & \bf bPQ $\uparrow$ \\
    \midrule
    ResNet50\cite{koonce2021resnet}+U-Net*\cite{ronneberger2015u}     &51.34 & 76.1 &48.70 & 67.80 \\
    ResNet50\cite{koonce2021resnet}+\textbf{Ours}    & 69.81 &131.8 & 49.53 \textcolor{red}{(+0.83)} &68.33 \textcolor{red}{(+0.53)} \\
    ResNet50\cite{koonce2021resnet}+\textbf{Ours\textcolor{gray}{(Multi-decoders)}} &198.67 &348.4 &48.43 \textcolor{green}{(-1.10)} &67.77 \textcolor{green}{(-0.56)} \\
    \midrule
    ViT-S\cite{touvron2021training}+U-Net*\cite{ronneberger2015u}  &104.42 &153.9 &26.94&43.28 \\
    ViT-S\cite{touvron2021training}+\textbf{Ours}  &136.33 & 153.9&35.93 \textcolor{red}{(+8.99)} &54.47 \textcolor{red}{(+11.19)} \\
    ViT-S\cite{touvron2021training}+\textbf{Ours\textcolor{gray}{(Multi-decoders)}}&398.67&258.2&17.16 \textcolor{green}{(-18.77)} &32.07 \textcolor{green}{(-22.40)} \\
    \midrule
    \textbf {LKNet-B}~\cite{ding2023unireplknet}+U-Net*\cite{ronneberger2015u}     & 39.88 &153.9 &50.13 & 68.11 \\
    \textbf {LKNet-B}~\cite{ding2023unireplknet}+\textbf{Ours}  &46.25 &163.8 &50.50 \textcolor{red}{(+0.37)} &68.51 \textcolor{red}{(+0.40)}\\
    \textbf {LKNet-B}~\cite{ding2023unireplknet}+\textbf{Ours\textcolor{gray}{(Multi-decoders)}} &134.56 & 268.4 & 49.89 \textcolor{green}{(-0.61)} & 68.19 \textcolor{green}{(-0.32)}\\
\bottomrule
\end{tabular}}
\end{table}